\documentclass[prl,twocolumn,superscriptaddress]{revtex4-2}
\usepackage[colorlinks=true,linkcolor=blue,citecolor=red,plainpages=false,pdfpagelabels]{hyperref}
\usepackage{mathtools}
\usepackage{graphicx}
\usepackage{amsmath}
\usepackage{braket}
\usepackage{cleveref}
\usepackage{amssymb}

\newcommand{\cX}{\mathcal{X}}
\usepackage{lipsum}  
\usepackage{xr}

\begin{document}

\title{Experimental Guesswork with Quantum Side Information using Twisted Light}
\author{Vishal Katariya}
\email{vkatar2@lsu.edu}
\affiliation{Hearne Institute for Theoretical Physics, Department of Physics and Astronomy, and Center for Computation and Technology, Louisiana State University, Baton Rouge, Louisiana 70803, USA}
\author{Narayan Bhusal} 
\affiliation{Quantum Photonics Laboratory, Department of Physics \& Astronomy, Louisiana State University, Baton Rouge, LA 70803, USA}
\author{Chenglong You}
\affiliation{Quantum Photonics Laboratory, Department of Physics \& Astronomy, Louisiana State University, Baton Rouge, LA 70803, USA}

\begin{abstract}
The guesswork is an information-theoretic quantity which can be seen as an alternate security criterion to entropy. Recent work has established the theoretical framework for guesswork in the presence of quantum side information, which we extend both theoretically and experimentally. We consider guesswork when the side information consists of the BB84 states and their higher-dimensional generalizations. With this side information, we compute the guesswork for two different scenarios for each dimension. We then perform a proof-of-principle experiment using Laguerre-Gauss modes to experimentally compute the guesswork for higher-dimensional generalizations of the BB84 states. We find that our experimental results agree closely with our theoretical predictions. This work shows that guesswork can be a viable security criterion in cryptographic tasks, and is experimentally accessible across a number of optical setups.
\end{abstract}

\maketitle

\section{Introduction}

Guesswork is an information-theoretic measure of security and uncertainty of an information source \cite{Massey1994,Arikan1996}, similar to entropy. In its simplest form, it can be understood as a game between two agents Alice and Bob. Alice picks an element $x$ from an alphabet $\cX$ with prior probability $p_X(x)$. Bob's task then is to guess Alice's choice $x$ while being allowed to ask questions of the form, ``Is $X=x$?" The guesswork, $G(X)$, is the average number of guesses Bob needs until Alice answers with, ``yes". This is in contrast to the entropy, where the same game is played except that Bob is allowed to ask questions of the form, ``Is $X \in \widetilde{X}?"$, where $\widetilde{X}$ is a subset of the alphabet $\cX$~\cite{book1991cover}. 

Guesswork also has real-world applicability. Consider that one's account on an online portal is subjected to a brute-force hacking attack. A malicious agent is only allowed a certain number of guesses at the password before being locked out. The average number of guesses, i.e. the guesswork, would be the operational criterion of security in such a situation. Furthermore, guesswork takes on richer behavior when Bob possesses some quantum correlations with Alice, also known as side information. The theoretical framework for this problem has been laid out and studied recently \cite{CCWF15,Hanson2021}. However, experimental verification of the guesswork with quantum side information is yet an unexplored avenue. 

Recently, spatially structured beams of light have been used extensively for multiple applications, such as 3D surface imaging, quantum cryptography, remote sensing and correlated imaging~\cite{bell:1999,geng:2010,lavery:2013,malik:2014,chen:2014,mirho:2015,omar:2016,dunlop:2016, yang2017,milione:2017,omarm:2019,jack:2009}. Among them, Laguerre-Gauss (LG) modes are particularly important as they possess orbital angular momentum (OAM)~\cite{siegman:1986,allen1992orbital} and allow for the construction of OAM modes of light. OAM modes enable the construction of orthonormal bases of light in any arbitrary finite dimension. OAM also enables the construction of the mutually unbiased basis of azimuthal angle (ANG) \cite{Giovannini2013,DAmbrosio2013,Malik12}. These two properties allow for the generation of the qubit BB84 states \cite{BB84} as well as their higher dimensional generalizations, which are especially important in quantum cryptography.

In this work, we perform a proof-of-principle experiment where we use spatial modes of light to experimentally calculate and verify the value of guesswork for several physically and cryptographically relevant examples involving the BB84 states. We extend the work of Hanson \cite{Hanson2021}, both theoretically and experimentally, to higher dimensional generalizations. We find excellent agreement between our experimental results and theoretical predictions. In each of the cases considered, we find that there is a ``quantum'' gap in the guesswork between the standard basis measurement and the optimal projective measurement. Our work shows that guesswork can be a viable security criterion in cryptographic tasks, and is experimentally accessible from optical setups.

\section{Theory} \label{sec:guesswork}

\begin{figure}[!hbtp]
	\includegraphics[width=\linewidth]{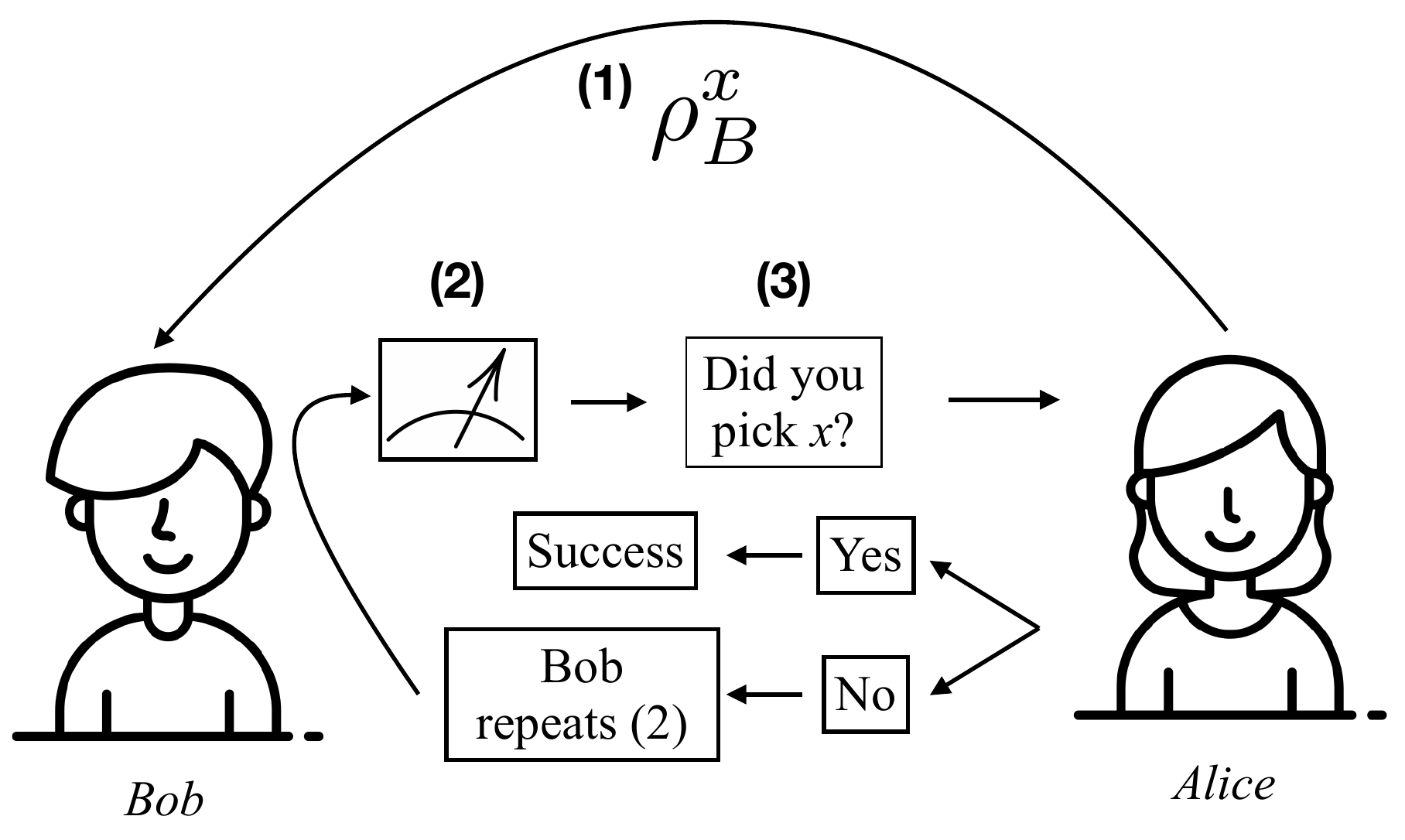}
	\caption{Guesswork can be understood via this guessing game played by Alice and Bob. Alice picks a classical symbol $x$ and sends Bob the corresponding quantum state. Bob guesses Alice's symbol with the help of his quantum state $\rho_B^x$. In each round of the game, Bob performs a quantum measurement and uses the classical outcome to make a guess. He repeats this process until he guesses correctly.}
	\label{fig:guessing-game}
\end{figure}

First, we introduce the theoretical framework of guesswork with quantum side information. Guesswork with quantum side information can be viewed as a multi-round two-party game, as shown in Fig. \ref{fig:guessing-game}. The guesser, Bob, possesses a classical system, or more generally, a quantum system $B$, which is correlated with Alice's random variable $X$. In this work, we consider the latter (and more general case) case where Bob possesses quantum side information. This scenario is fully characterized by a classical-quantum state $\rho_{XB}$ shared by Alice and Bob, where Alice's symbols and their associated probabilities are captured in classical register $X$ and Bob's quantum side information is present in quantum system $B$. In the guessing game picture of guesswork, Alice picks an element $x \in \mathcal{X}$ and sends Bob the corresponding quantum state $\rho_B^x$. Bob then performs quantum measurements on his side information state $\rho_B^x$ whose results inform his guessing strategy. In general, Bob performs a quantum instrument before each guess. A quantum instrument yields both a classical measurement outcome and a post-measurement quantum state. This ensures that after each round, Bob has classical information used to make a guess, and also a quantum state for future rounds of the guessing game. 

It was shown in Ref.~\cite{Hanson2021}, however, that Bob can do just as well if he performs a single quantum measurement to decide his guessing order. That is, the typical sequential guessing strategy can be reduced to a single-round guessing strategy. This divides the guessing game into two parts: an initial step involving a quantum measurement, followed by a purely classical guessing game between Alice and Bob. Such an equivalence makes a trade-off between time and space, in the sense that Bob needs more spatial resources and fewer temporal resources than the sequential guessing game.

A simple and yet instructive example of guesswork with quantum side information is that involving the four BB84 states \cite{BB84}. In this case, Alice first picks one of four classical letters $x_1$ through $x_4$ with equal probability, and then sends the corresponding BB84 states $\{\ket{0}, \ket{1}, \ket{+}, \ket{-} \}$ to Bob. Therefore, Bob's side information is hidden in his quantum state. Bob's task is to use his received quantum state to guess which classical letter Alice chose. Suppose that the projective measurement is characterized by the two orthogonal states $\{\ket{\psi(\theta)}, \ket{\psi(\pi/2-\theta)} \}$ where $\ket{\psi(\theta)} := \cos \theta \ket{0} + \sin \theta \ket{1}$. If Bob simply measures in the standard basis $\{\ket{\psi(0)}, \ket{\psi(\pi/2)} \}$, the average number of guesses is $1.75$ (See Supplementary Material). However, there exists a projective measurement which leads to a smaller guesswork. This optimal measurement is characterized by $\theta = 1/2 \arctan (1/3)$. This measurement can be shown to attain a guesswork of $1.709$~\cite{Hanson2021}.

From the above simple example, we see a clear separation of the guesswork when using the optimal projective measurement as compared to a standard basis measurement. Such a ``quantum'' separation of the guesswork can also be obtained in higher dimensional generalizations of the BB84 example. That is, we consider the side information system in the $d$-dimension BB84 generalization. Alice will pick one of $2d$ classical symbols with equal probability, each of which is correlated with one of the $2d$ side information states. These $2d$ states are divided into two mutually unbiased bases of $d$ states each. The states are as follows: 
\begin{equation}
	\{ \ket{0}, \ket{1}, \dots, \ket{d-1}, \ket{\widetilde{0}}, \ket{\widetilde{1}}, \dots, \ket{\widetilde{d-1}} \}
\end{equation}
where $\ket{\widetilde{j}} = \frac{1}{\sqrt{d}} \sum_{k=0}^{d-1} e^{i 2 \pi  k j/d} \ket{k} $ and $| \braket{i | \widetilde{j}} |^2 = 1/d ~\forall~ i, j \in \{ 0, 1, \dots, d-1 \}$. In this case, a standard basis measurement by Bob means that he projects his state onto the basis $\{ \ket{0}, \ket{1}, \dots, \ket{d-1} \}$. When outcome $\ket{k}$ is obtained, Bob can eliminate the $d-1$ standard basis states that are orthogonal to $\ket{k}$. His best strategy, in this case, is then to guess outcome $k$ first, then $\widetilde{0}$ through $\widetilde{d-1}$, and finally the remaining labels in any order. The guesswork in this case is $(d+5)/4$. We provide more details of this calculation in the Supplementary Material.

However, like the 2-dimensional case, Bob can do better by carefully selecting his quantum measurement. We briefly explain the strategy he can use and how it can be optimized. Consider that Bob chooses to project onto an arbitrarily chosen orthonormal basis $\{ \ket{\psi_0}, \dots, \ket{\psi_{d-1}} \}$. If he obtains the outcome corresponding to $\ket{\psi_k}$, then he guesses in decreasing order of the overlap between $\ket{\psi_k}$ and the $2d$ input states. We note again here that the post-measurement guessing strategy is purely classical, and we obtain it by invoking Massey's observation~\cite{Massey1994} to minimize the guesswork by guessing in decreasing order of the posterior probability of classical symbols \footnote{We note here that in general, a projective measurement does not achieve the minimum possible guesswork. To do so, one would need to optimize over all quantum measurements, not just projective ones.}. 

Since the rules of the game are decided beforehand, Bob finds and decides on his optimal projective measurement via a numerical technique. We perform this optimization for dimensions $d=3$ and $4$ using MATLAB. This optimization also yields the guessing order to use with each of these measurements. Using this technique, we find that there is a significant gap between the guesswork attained by standard basis measurements and that attained by the optimized projective measurement. We summarize this gap in Table~\ref{table:results} and provide more details in the Supplementary Material. 

\section{Experiment} 

\label{sec:experiment-details}

We now proceed to describe the experimental apparatus and techniques used to perform our experiment. The generalized BB84 states consist of two mutually unbiased bases. The bases we use in our experiment are the OAM basis and the ANG basis, which are mutually unbiased. Each mode of light, in the OAM basis, is characterized by an angular momentum quantum number $\ell$. In principle, $\ell$ can be any integer, and therefore the OAM basis consists of an infinite number of orthogonal modes. Thus, it is simple to generate an orthonormal basis in arbitrary dimension by selecting the appropriate values of $\ell$.

For example, if Alice generates the OAM basis consisting of qunatum numbers $\ell\in\{ -L,-L+1,...,L-1,L\}$, then we have a $d=2L+1$-dimensional OAM basis, consisting of the states
$\left\{ \Psi_{\text{OAM}}^\ell = e^{i \ell \varphi} \right\}_{\ell=-L}^{\ell=L}$.

The ANG basis corresponding to the OAM basis defined above also consists of $2L+1$ orthogonal states. Each ANG state is a superposition of each of the OAM states. The basis is constituted by the following states:
\begin{equation}
		\left\{ \Psi_{\text{ANG}}^n = \frac{1}{\sqrt{d}} \sum_{\ell=-L}^{\ell=L} e^{\frac{2 i \pi n \ell}{d}} \Psi_{\text{OAM}}^\ell  \right\}_{n=-L}^{n=L}.
\end{equation}

Simple verification shows that the two bases are indeed mutually unbiased, i.e.,
\begin{equation}
	| \langle \Psi_{\text{OAM}}^l | \Psi_{\text{ANG}}^n \rangle|^2 = 1/d.
\end{equation}

\subsection{Experimental setup}

\begin{figure}[!htbp]
	\includegraphics[width=0.85\linewidth]{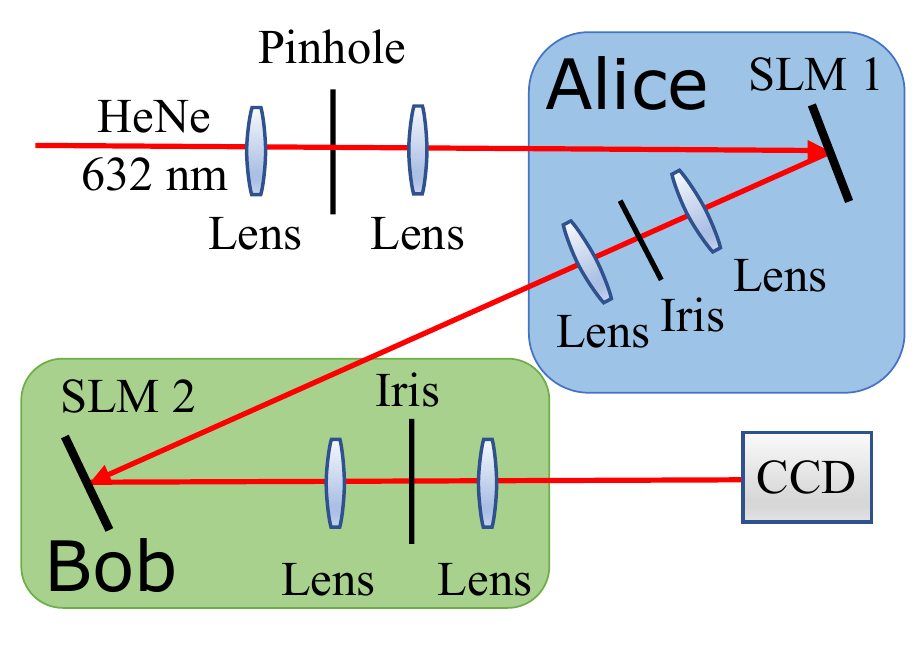} 
	\caption{The schematic diagram of the setup used to demonstrate guesswork with quantum side information. The experiment utilizes a He-Ne laser whose output spatial mode is first cleaned. The higher-dimensional generalizations of the BB84 states are prepared by Alice using a spatial light modulator (SLM). The prepared states are sent to Bob through a free-space communication channel, a 4f system. Bob then performs his quantum measurement using a second SLM and a charged coupled device (CCD) camera.}
	\label{fig:experimental-setup}
\end{figure}

The schematic diagram of our experimental setup is depicted in Figure~\ref{fig:experimental-setup}. Here, Alice prepares quantum states of light which corresponds to her choice of symbol, and then sends to Bob for further processing. In our experiment, Alice use a spatial light modulator (SLM) and computer-generated holograms to generate the LG modes required in our experiment~\cite{Ando09}. This technique enables us to generate spatial modes in the first-order diffraction order of the SLM. This is sufficient to generate the OAM and ANG modes that correspond to generalized BB84 states in higher dimensions. The generated modes are filtered and sent to Bob using a 4f-optical system. Bob then projects the spatial mode onto a second SLM to perform a projective measurement. The specific holograms imprinted on the SLM dictate the basis onto which the initial mode is projected. The beam reflected by the second SLM corresponds to a post-measurement state, which is propagated to a charge-coupled device (CCD). The spatial profile of the final beam is captured and analysed. This process is repeated for each of the measurement states to be projected on, and the captured images are then used to compute the guesswork.

\subsection{Experimental Determination of Guesswork} 

Our goal is to compute the guesswork for the generalized BB84 states for dimensions $d=2, 3$, and $4$. For each dimension, the guesswork is computed for the standard basis measurement as well as the optimized projective measurement, which makes for a total of six different scenarios.

In each iteration of the guessing game, Bob begins with a predecided guessing order for each possible measurement outcome. For each state Alice sends, he projects onto each of the basis states that characterize his measurement. This enables him to determine the relative rate of the measurement outcomes and hence decide on the measurement outcome.
This holds for both the standard and optimized basis measurement. Once the guessing order is decided, Bob simply sends Alice his guesses--this interaction yields the average number of guesses for each input state. Averaging over input states yields the overall guesswork. The post-processing of CCD images to compute the guesswork is done using MATLAB.

We perform a total of six experiments corresponding to six scenarios: this comprises two sets of measurements for each of the three dimensions considered. The two measurements are the standard basis measurement and the numerically determined optimal projective measurement. In each case, we repeated the measurement ten times for each input state. Overall, this is equivalent to playing the guesswork game ten times for each of the six scenarios under consideration.

In Table~\ref{table:results}, we present our experimental results. The results are divided into two parts: the guesswork when performing a standard basis measurement, and when performing the optimized projective measurement. We see that for all scenarios we consider, the guesswork is within $1\%$ of the theoretical prediction for both the standard basis and optimal measurement.

\small
\begin{table}[!t]
\centering
\caption{The guesswork for each of the scenarios considered.}
\label{table:results}
{\def\arraystretch{1}\tabcolsep=8pt
\begin{tabular}{ccc}
	\hline 
		 
		 Dimension & Theoretical value & Experimental value \\
		 \hline
		 \multicolumn{3}{c}{Standard basis measurement} \\
d=2 & 1.75        & $1.7505 \pm 0.0017$ \\
d=3 & 2           & $1.9996 \pm 0.0087$ \\
d=4 & 2.25        & $2.2547 \pm 0.0029$ \\
\multicolumn{3}{c}{Optimized projective measurement} \\
d=2 & 1.709     & $1.7062 \pm 0.0089$ \\
d=3 & 1.9425    & $1.9439 \pm 0.0084$ \\
d=4 & 2.1429    & $2.1411 \pm 0.0025$ \\
\hline
\end{tabular}
}
\end{table}
\normalsize

\begin{figure}[!ht]
	\includegraphics[width=0.85\linewidth]{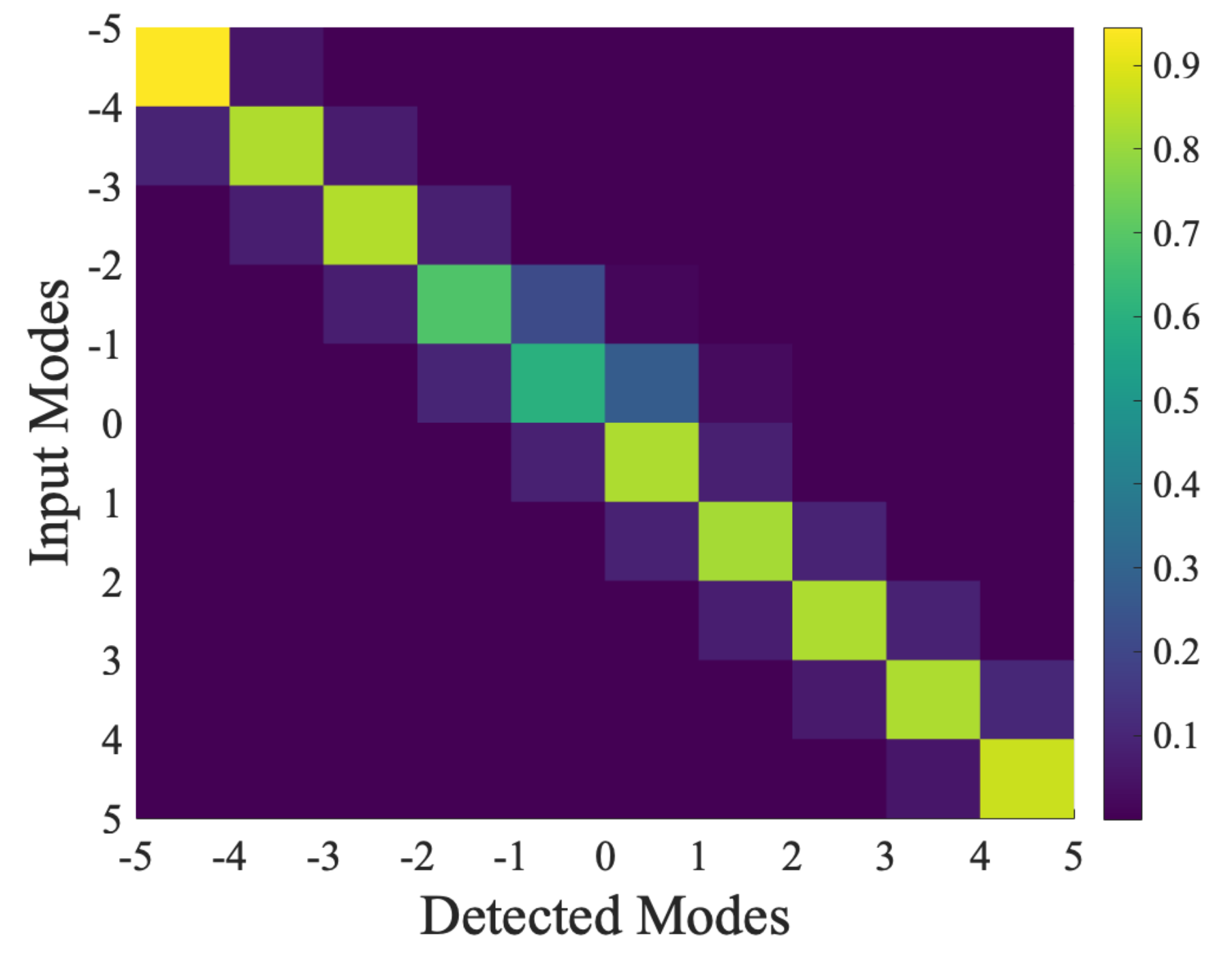} 
	\caption{The cross-correlation matrix representing the conditional probabilities between sent and detected modes in the OAM basis. The off-diagonal elements in the figure indicate cross-talks between adjacent modes. The experimental overlap of two adjacent OAM modes is small, indicating a good selection of OAM modes for the experiment.}
	\label{fig:overlap-table}
\end{figure}

Finally, we remark on the choice of specific OAM modes used for each dimension $d$. OAM modes are, in principle, orthogonal to each other. Thus any combination of them can be used to construct the desired orthonormal basis. However, due to the finite size of the SLM pixels, they are not perfectly orthogonal in practice. We quantify the overlap between OAM modes under consideration using a cross-correlation matrix shown in Figure~\ref{fig:overlap-table}. To minimize overlap errors in our experiment, we choose $\ell$ values that are spaced further apart from one another to generate our input states, instead of using consecutive $\ell$ values. For $d=2$, we use the $\ell$ values $\{-3,3\}$. Similarly for $d=3$ and $d=4$, we use $\ell$ values $\{-3,0,3\}$ and $\{-3,-1,1,3\}$ respectively. We note that this modification does not alter any of the calculations for the guesswork itself, and is done only to reduce errors that arise from non-zero overlap between nearby OAM modes. 

\section{Conclusion}

In summary, we showed how to use accessible spatial modes of light to experimentally compute the guesswork in the presence of side information. We considered the side information to be higher-dimensional generalizations of the qubit BB84 states, and showed that the experimentally calculated guesswork matches theoretical predictions to within an error of $1\%$. This proof-of-principle work lays the ground for further experimental applications of guesswork with quantum side information. We hope that this work will serve as a starting point for more experimental uses of the guesswork as a security criterion.

\section{Acknowledgements}

The authors thank O.S.M.L. for the helpful discussions. V.K. acknowledges support from the LSU Economic Development Assistantship and the US National Science Foundation via grant number 1907615. C.Y. acknowledge funding from the U.S. Department of Energy, Office of Basic Energy Sciences, Division of Materials Sciences and Engineering under Award DE-SC0021069.

\bibliography{guesswork-exp-refs}

\begin{thebibliography}{25}%
\makeatletter
\providecommand \@ifxundefined [1]{%
 \@ifx{#1\undefined}
}%
\providecommand \@ifnum [1]{%
 \ifnum #1\expandafter \@firstoftwo
 \else \expandafter \@secondoftwo
 \fi
}%
\providecommand \@ifx [1]{%
 \ifx #1\expandafter \@firstoftwo
 \else \expandafter \@secondoftwo
 \fi
}%
\providecommand \natexlab [1]{#1}%
\providecommand \enquote  [1]{``#1''}%
\providecommand \bibnamefont  [1]{#1}%
\providecommand \bibfnamefont [1]{#1}%
\providecommand \citenamefont [1]{#1}%
\providecommand \href@noop [0]{\@secondoftwo}%
\providecommand \href [0]{\begingroup \@sanitize@url \@href}%
\providecommand \@href[1]{\@@startlink{#1}\@@href}%
\providecommand \@@href[1]{\endgroup#1\@@endlink}%
\providecommand \@sanitize@url [0]{\catcode `\\12\catcode `\$12\catcode
  `\&12\catcode `\#12\catcode `\^12\catcode `\_12\catcode `\%12\relax}%
\providecommand \@@startlink[1]{}%
\providecommand \@@endlink[0]{}%
\providecommand \url  [0]{\begingroup\@sanitize@url \@url }%
\providecommand \@url [1]{\endgroup\@href {#1}{\urlprefix }}%
\providecommand \urlprefix  [0]{URL }%
\providecommand \Eprint [0]{\href }%
\providecommand \doibase [0]{https://doi.org/}%
\providecommand \selectlanguage [0]{\@gobble}%
\providecommand \bibinfo  [0]{\@secondoftwo}%
\providecommand \bibfield  [0]{\@secondoftwo}%
\providecommand \translation [1]{[#1]}%
\providecommand \BibitemOpen [0]{}%
\providecommand \bibitemStop [0]{}%
\providecommand \bibitemNoStop [0]{.\EOS\space}%
\providecommand \EOS [0]{\spacefactor3000\relax}%
\providecommand \BibitemShut  [1]{\csname bibitem#1\endcsname}%
\let\auto@bib@innerbib\@empty
\bibitem [{\citenamefont {Massey}(1994)}]{Massey1994}%
  \BibitemOpen
  \bibfield  {author} {\bibinfo {author} {\bibfnamefont {J.}~\bibnamefont
  {Massey}},\ }\bibfield  {title} {\bibinfo {title} {Guessing and entropy},\
  }in\ \href {https://doi.org/10.1109/ISIT.1994.394764} {\emph {\bibinfo
  {booktitle} {Proceedings of 1994 IEEE International Symposium on Information
  Theory}}}\ (\bibinfo {year} {1994})\ p.\ \bibinfo {pages} {204}\BibitemShut
  {NoStop}%
\bibitem [{\citenamefont {Arikan}(1996)}]{Arikan1996}%
  \BibitemOpen
  \bibfield  {author} {\bibinfo {author} {\bibfnamefont {E.}~\bibnamefont
  {Arikan}},\ }\bibfield  {title} {\bibinfo {title} {An inequality on guessing
  and its application to sequential decoding},\ }\href
  {https://doi.org/10.1109/18.481781} {\bibfield  {journal} {\bibinfo
  {journal} {IEEE Transactions on Information Theory}\ }\textbf {\bibinfo
  {volume} {42}},\ \bibinfo {pages} {99} (\bibinfo {year} {1996})}\BibitemShut
  {NoStop}%
\bibitem [{\citenamefont {Cover}\ and\ \citenamefont
  {Thomas}(2006)}]{book1991cover}%
  \BibitemOpen
  \bibfield  {author} {\bibinfo {author} {\bibfnamefont {T.~M.}\ \bibnamefont
  {Cover}}\ and\ \bibinfo {author} {\bibfnamefont {J.~A.}\ \bibnamefont
  {Thomas}},\ }\href@noop {} {\emph {\bibinfo {title} {Elements of Information
  Theory}}},\ \bibinfo {edition} {2nd}\ ed.\ (\bibinfo  {publisher}
  {Wiley-Interscience},\ \bibinfo {year} {2006})\BibitemShut {NoStop}%
\bibitem [{\citenamefont {Chen}\ \emph {et~al.}(2015)\citenamefont {Chen},
  \citenamefont {Cao}, \citenamefont {Wang},\ and\ \citenamefont
  {Feng}}]{CCWF15}%
  \BibitemOpen
  \bibfield  {author} {\bibinfo {author} {\bibfnamefont {W.}~\bibnamefont
  {Chen}}, \bibinfo {author} {\bibfnamefont {Y.}~\bibnamefont {Cao}}, \bibinfo
  {author} {\bibfnamefont {H.}~\bibnamefont {Wang}},\ and\ \bibinfo {author}
  {\bibfnamefont {Y.}~\bibnamefont {Feng}},\ }\bibfield  {title} {\bibinfo
  {title} {Minimum guesswork discrimination between quantum states},\
  }\href@noop {} {\bibfield  {journal} {\bibinfo  {journal} {Quantum Info.
  Comput.}\ }\textbf {\bibinfo {volume} {15}},\ \bibinfo {pages} {737–758}
  (\bibinfo {year} {2015})}\BibitemShut {NoStop}%
\bibitem [{\citenamefont {Hanson}\ \emph {et~al.}(2022)\citenamefont {Hanson},
  \citenamefont {Katariya}, \citenamefont {Datta},\ and\ \citenamefont
  {Wilde}}]{Hanson2021}%
  \BibitemOpen
  \bibfield  {author} {\bibinfo {author} {\bibfnamefont {E.~P.}\ \bibnamefont
  {Hanson}}, \bibinfo {author} {\bibfnamefont {V.}~\bibnamefont {Katariya}},
  \bibinfo {author} {\bibfnamefont {N.}~\bibnamefont {Datta}},\ and\ \bibinfo
  {author} {\bibfnamefont {M.~M.}\ \bibnamefont {Wilde}},\ }\bibfield  {title}
  {\bibinfo {title} {Guesswork with quantum side information},\ }\href
  {https://doi.org/10.1109/TIT.2021.3118878} {\bibfield  {journal} {\bibinfo
  {journal} {IEEE Transactions on Information Theory}\ }\textbf {\bibinfo
  {volume} {68}},\ \bibinfo {pages} {322} (\bibinfo {year} {2022})}\BibitemShut
  {NoStop}%
\bibitem [{\citenamefont {Bell}\ \emph {et~al.}(1999)\citenamefont {Bell},
  \citenamefont {Li},\ and\ \citenamefont {Zhang}}]{bell:1999}%
  \BibitemOpen
  \bibfield  {author} {\bibinfo {author} {\bibfnamefont {T.}~\bibnamefont
  {Bell}}, \bibinfo {author} {\bibfnamefont {B.}~\bibnamefont {Li}},\ and\
  \bibinfo {author} {\bibfnamefont {S.}~\bibnamefont {Zhang}},\ }\bibfield
  {title} {\bibinfo {title} {Structured light techniques and applications},\
  }\href@noop {} {\bibfield  {journal} {\bibinfo  {journal} {Wiley Encyclopedia
  of Electrical and Electronics Engineering}\ ,\ \bibinfo {pages} {1}}
  (\bibinfo {year} {1999})}\BibitemShut {NoStop}%
\bibitem [{\citenamefont {Geng}(2011)}]{geng:2010}%
  \BibitemOpen
  \bibfield  {author} {\bibinfo {author} {\bibfnamefont {J.}~\bibnamefont
  {Geng}},\ }\bibfield  {title} {\bibinfo {title} {Structured-light 3d surface
  imaging: a tutorial},\ }\href@noop {} {\bibfield  {journal} {\bibinfo
  {journal} {Adv. Opt. Photonics}\ }\textbf {\bibinfo {volume} {3}},\ \bibinfo
  {pages} {128} (\bibinfo {year} {2011})}\BibitemShut {NoStop}%
\bibitem [{\citenamefont {Lavery}\ \emph {et~al.}(2013)\citenamefont {Lavery},
  \citenamefont {Speirits}, \citenamefont {Barnett},\ and\ \citenamefont
  {Padgett}}]{lavery:2013}%
  \BibitemOpen
  \bibfield  {author} {\bibinfo {author} {\bibfnamefont {M.~P.}\ \bibnamefont
  {Lavery}}, \bibinfo {author} {\bibfnamefont {F.~C.}\ \bibnamefont
  {Speirits}}, \bibinfo {author} {\bibfnamefont {S.~M.}\ \bibnamefont
  {Barnett}},\ and\ \bibinfo {author} {\bibfnamefont {M.~J.}\ \bibnamefont
  {Padgett}},\ }\bibfield  {title} {\bibinfo {title} {Detection of a spinning
  object using light’s orbital angular momentum},\ }\href@noop {} {\bibfield
  {journal} {\bibinfo  {journal} {Science}\ }\textbf {\bibinfo {volume}
  {341}},\ \bibinfo {pages} {537} (\bibinfo {year} {2013})}\BibitemShut
  {NoStop}%
\bibitem [{\citenamefont {Malik}\ and\ \citenamefont
  {Boyd}(2014)}]{malik:2014}%
  \BibitemOpen
  \bibfield  {author} {\bibinfo {author} {\bibfnamefont {M.}~\bibnamefont
  {Malik}}\ and\ \bibinfo {author} {\bibfnamefont {R.}~\bibnamefont {Boyd}},\
  }\bibfield  {title} {\bibinfo {title} {Quantum imaging technologies},\
  }\href@noop {} {\bibfield  {journal} {\bibinfo  {journal} {RIVISTA DEL NUOVO
  CIMENTO}\ }\textbf {\bibinfo {volume} {37}} (\bibinfo {year}
  {2014})}\BibitemShut {NoStop}%
\bibitem [{\citenamefont {Chen}\ \emph {et~al.}(2014)\citenamefont {Chen},
  \citenamefont {Lei},\ and\ \citenamefont {Romero}}]{chen:2014}%
  \BibitemOpen
  \bibfield  {author} {\bibinfo {author} {\bibfnamefont {L.}~\bibnamefont
  {Chen}}, \bibinfo {author} {\bibfnamefont {J.}~\bibnamefont {Lei}},\ and\
  \bibinfo {author} {\bibfnamefont {J.}~\bibnamefont {Romero}},\ }\bibfield
  {title} {\bibinfo {title} {Quantum digital spiral imaging},\ }\href@noop {}
  {\bibfield  {journal} {\bibinfo  {journal} {Light: Science \& Applications}\
  }\textbf {\bibinfo {volume} {3}},\ \bibinfo {pages} {e153} (\bibinfo {year}
  {2014})}\BibitemShut {NoStop}%
\bibitem [{\citenamefont {Mirhosseini}\ \emph {et~al.}(2015)\citenamefont
  {Mirhosseini}, \citenamefont {Maga{\~n}a-Loaiza}, \citenamefont
  {O’Sullivan}, \citenamefont {Rodenburg}, \citenamefont {Malik},
  \citenamefont {Lavery}, \citenamefont {Padgett}, \citenamefont {Gauthier},\
  and\ \citenamefont {Boyd}}]{mirho:2015}%
  \BibitemOpen
  \bibfield  {author} {\bibinfo {author} {\bibfnamefont {M.}~\bibnamefont
  {Mirhosseini}}, \bibinfo {author} {\bibfnamefont {O.~S.}\ \bibnamefont
  {Maga{\~n}a-Loaiza}}, \bibinfo {author} {\bibfnamefont {M.~N.}\ \bibnamefont
  {O’Sullivan}}, \bibinfo {author} {\bibfnamefont {B.}~\bibnamefont
  {Rodenburg}}, \bibinfo {author} {\bibfnamefont {M.}~\bibnamefont {Malik}},
  \bibinfo {author} {\bibfnamefont {M.~P.}\ \bibnamefont {Lavery}}, \bibinfo
  {author} {\bibfnamefont {M.~J.}\ \bibnamefont {Padgett}}, \bibinfo {author}
  {\bibfnamefont {D.~J.}\ \bibnamefont {Gauthier}},\ and\ \bibinfo {author}
  {\bibfnamefont {R.~W.}\ \bibnamefont {Boyd}},\ }\bibfield  {title} {\bibinfo
  {title} {High-dimensional quantum cryptography with twisted light},\
  }\href@noop {} {\bibfield  {journal} {\bibinfo  {journal} {New J. Phys.}\
  }\textbf {\bibinfo {volume} {17}},\ \bibinfo {pages} {033033} (\bibinfo
  {year} {2015})}\BibitemShut {NoStop}%
\bibitem [{\citenamefont {Maga{\~n}a-Loaiza}\ \emph {et~al.}(2016)\citenamefont
  {Maga{\~n}a-Loaiza}, \citenamefont {Mirhosseini}, \citenamefont {Cross},
  \citenamefont {Rafsanjani},\ and\ \citenamefont {Boyd}}]{omar:2016}%
  \BibitemOpen
  \bibfield  {author} {\bibinfo {author} {\bibfnamefont {O.~S.}\ \bibnamefont
  {Maga{\~n}a-Loaiza}}, \bibinfo {author} {\bibfnamefont {M.}~\bibnamefont
  {Mirhosseini}}, \bibinfo {author} {\bibfnamefont {R.~M.}\ \bibnamefont
  {Cross}}, \bibinfo {author} {\bibfnamefont {S.~M.~H.}\ \bibnamefont
  {Rafsanjani}},\ and\ \bibinfo {author} {\bibfnamefont {R.~W.}\ \bibnamefont
  {Boyd}},\ }\bibfield  {title} {\bibinfo {title} {Hanbury brown and twiss
  interferometry with twisted light},\ }\href@noop {} {\bibfield  {journal}
  {\bibinfo  {journal} {Sci. Adv.}\ }\textbf {\bibinfo {volume} {2}},\ \bibinfo
  {pages} {e1501143} (\bibinfo {year} {2016})}\BibitemShut {NoStop}%
\bibitem [{\citenamefont {Rubinsztein-Dunlop}\ \emph
  {et~al.}(2016)\citenamefont {Rubinsztein-Dunlop}, \citenamefont {Forbes},
  \citenamefont {Berry}, \citenamefont {Dennis}, \citenamefont {Andrews},
  \citenamefont {Mansuripur}, \citenamefont {Denz}, \citenamefont {Alpmann},
  \citenamefont {Banzer}, \citenamefont {Bauer} \emph {et~al.}}]{dunlop:2016}%
  \BibitemOpen
  \bibfield  {author} {\bibinfo {author} {\bibfnamefont {H.}~\bibnamefont
  {Rubinsztein-Dunlop}}, \bibinfo {author} {\bibfnamefont {A.}~\bibnamefont
  {Forbes}}, \bibinfo {author} {\bibfnamefont {M.~V.}\ \bibnamefont {Berry}},
  \bibinfo {author} {\bibfnamefont {M.~R.}\ \bibnamefont {Dennis}}, \bibinfo
  {author} {\bibfnamefont {D.~L.}\ \bibnamefont {Andrews}}, \bibinfo {author}
  {\bibfnamefont {M.}~\bibnamefont {Mansuripur}}, \bibinfo {author}
  {\bibfnamefont {C.}~\bibnamefont {Denz}}, \bibinfo {author} {\bibfnamefont
  {C.}~\bibnamefont {Alpmann}}, \bibinfo {author} {\bibfnamefont
  {P.}~\bibnamefont {Banzer}}, \bibinfo {author} {\bibfnamefont
  {T.}~\bibnamefont {Bauer}}, \emph {et~al.},\ }\bibfield  {title} {\bibinfo
  {title} {Roadmap on structured light},\ }\href@noop {} {\bibfield  {journal}
  {\bibinfo  {journal} {J. Opt.}\ }\textbf {\bibinfo {volume} {19}},\ \bibinfo
  {pages} {013001} (\bibinfo {year} {2016})}\BibitemShut {NoStop}%
\bibitem [{\citenamefont {Yang}\ \emph {et~al.}(2017)\citenamefont {Yang},
  \citenamefont {Maga{\~n}a-Loaiza}, \citenamefont {Mirhosseini}, \citenamefont
  {Zhou}, \citenamefont {Gao}, \citenamefont {Gao}, \citenamefont {Rafsanjani},
  \citenamefont {Long},\ and\ \citenamefont {Boyd}}]{yang2017}%
  \BibitemOpen
  \bibfield  {author} {\bibinfo {author} {\bibfnamefont {Z.}~\bibnamefont
  {Yang}}, \bibinfo {author} {\bibfnamefont {O.~S.}\ \bibnamefont
  {Maga{\~n}a-Loaiza}}, \bibinfo {author} {\bibfnamefont {M.}~\bibnamefont
  {Mirhosseini}}, \bibinfo {author} {\bibfnamefont {Y.}~\bibnamefont {Zhou}},
  \bibinfo {author} {\bibfnamefont {B.}~\bibnamefont {Gao}}, \bibinfo {author}
  {\bibfnamefont {L.}~\bibnamefont {Gao}}, \bibinfo {author} {\bibfnamefont
  {S.~M.~H.}\ \bibnamefont {Rafsanjani}}, \bibinfo {author} {\bibfnamefont
  {G.-L.}\ \bibnamefont {Long}},\ and\ \bibinfo {author} {\bibfnamefont
  {R.~W.}\ \bibnamefont {Boyd}},\ }\bibfield  {title} {\bibinfo {title}
  {Digital spiral object identification using random light},\ }\href@noop {}
  {\bibfield  {journal} {\bibinfo  {journal} {Light: Science \& Applications}\
  }\textbf {\bibinfo {volume} {6}},\ \bibinfo {pages} {e17013} (\bibinfo {year}
  {2017})}\BibitemShut {NoStop}%
\bibitem [{\citenamefont {Milione}\ \emph {et~al.}(2017)\citenamefont
  {Milione}, \citenamefont {Wang}, \citenamefont {Han},\ and\ \citenamefont
  {Bai}}]{milione:2017}%
  \BibitemOpen
  \bibfield  {author} {\bibinfo {author} {\bibfnamefont {G.}~\bibnamefont
  {Milione}}, \bibinfo {author} {\bibfnamefont {T.}~\bibnamefont {Wang}},
  \bibinfo {author} {\bibfnamefont {J.}~\bibnamefont {Han}},\ and\ \bibinfo
  {author} {\bibfnamefont {L.}~\bibnamefont {Bai}},\ }\bibfield  {title}
  {\bibinfo {title} {Remotely sensing an object’s rotational orientation
  using the orbital angular momentum of light},\ }\href@noop {} {\bibfield
  {journal} {\bibinfo  {journal} {Chin. Opt. Lett.}\ }\textbf {\bibinfo
  {volume} {15}},\ \bibinfo {pages} {030012} (\bibinfo {year}
  {2017})}\BibitemShut {NoStop}%
\bibitem [{\citenamefont {Maga{\~n}a-Loaiza}\ \emph {et~al.}(2019)\citenamefont
  {Maga{\~n}a-Loaiza}, \citenamefont {Le{\'o}n-Montiel}, \citenamefont
  {Perez-Leija}, \citenamefont {U'Ren}, \citenamefont {You}, \citenamefont
  {Busch}, \citenamefont {Lita}, \citenamefont {Nam}, \citenamefont {Mirin},\
  and\ \citenamefont {Gerrits}}]{omarm:2019}%
  \BibitemOpen
  \bibfield  {author} {\bibinfo {author} {\bibfnamefont {O.~S.}\ \bibnamefont
  {Maga{\~n}a-Loaiza}}, \bibinfo {author} {\bibfnamefont {R.~d.~J.}\
  \bibnamefont {Le{\'o}n-Montiel}}, \bibinfo {author} {\bibfnamefont
  {A.}~\bibnamefont {Perez-Leija}}, \bibinfo {author} {\bibfnamefont {A.~B.}\
  \bibnamefont {U'Ren}}, \bibinfo {author} {\bibfnamefont {C.}~\bibnamefont
  {You}}, \bibinfo {author} {\bibfnamefont {K.}~\bibnamefont {Busch}}, \bibinfo
  {author} {\bibfnamefont {A.~E.}\ \bibnamefont {Lita}}, \bibinfo {author}
  {\bibfnamefont {S.~W.}\ \bibnamefont {Nam}}, \bibinfo {author} {\bibfnamefont
  {R.~P.}\ \bibnamefont {Mirin}},\ and\ \bibinfo {author} {\bibfnamefont
  {T.}~\bibnamefont {Gerrits}},\ }\bibfield  {title} {\bibinfo {title}
  {Multiphoton quantum-state engineering using conditional measurements},\
  }\href@noop {} {\bibfield  {journal} {\bibinfo  {journal} {npj Quantum Inf.}\
  }\textbf {\bibinfo {volume} {5}},\ \bibinfo {pages} {80} (\bibinfo {year}
  {2019})}\BibitemShut {NoStop}%
\bibitem [{\citenamefont {Jack}\ \emph {et~al.}(2009)\citenamefont {Jack},
  \citenamefont {Leach}, \citenamefont {Romero}, \citenamefont {Franke-Arnold},
  \citenamefont {Ritsch-Marte}, \citenamefont {Barnett},\ and\ \citenamefont
  {Padgett}}]{jack:2009}%
  \BibitemOpen
  \bibfield  {author} {\bibinfo {author} {\bibfnamefont {B.}~\bibnamefont
  {Jack}}, \bibinfo {author} {\bibfnamefont {J.}~\bibnamefont {Leach}},
  \bibinfo {author} {\bibfnamefont {J.}~\bibnamefont {Romero}}, \bibinfo
  {author} {\bibfnamefont {S.}~\bibnamefont {Franke-Arnold}}, \bibinfo {author}
  {\bibfnamefont {M.}~\bibnamefont {Ritsch-Marte}}, \bibinfo {author}
  {\bibfnamefont {S.}~\bibnamefont {Barnett}},\ and\ \bibinfo {author}
  {\bibfnamefont {M.}~\bibnamefont {Padgett}},\ }\bibfield  {title} {\bibinfo
  {title} {Holographic ghost imaging and the violation of a bell inequality},\
  }\href@noop {} {\bibfield  {journal} {\bibinfo  {journal} {Phys. Rev. Lett.}\
  }\textbf {\bibinfo {volume} {103}},\ \bibinfo {pages} {083602} (\bibinfo
  {year} {2009})}\BibitemShut {NoStop}%
\bibitem [{\citenamefont {Siegman}(1986)}]{siegman:1986}%
  \BibitemOpen
  \bibfield  {author} {\bibinfo {author} {\bibfnamefont {A.~E.}\ \bibnamefont
  {Siegman}},\ }\href@noop {} {\emph {\bibinfo {title} {Lasers}}}\ (\bibinfo
  {publisher} {University Science Books},\ \bibinfo {year} {1986})\BibitemShut
  {NoStop}%
\bibitem [{\citenamefont {Allen}\ \emph {et~al.}(1992)\citenamefont {Allen},
  \citenamefont {Beijersbergen}, \citenamefont {Spreeuw},\ and\ \citenamefont
  {Woerdman}}]{allen1992orbital}%
  \BibitemOpen
  \bibfield  {author} {\bibinfo {author} {\bibfnamefont {L.}~\bibnamefont
  {Allen}}, \bibinfo {author} {\bibfnamefont {M.~W.}\ \bibnamefont
  {Beijersbergen}}, \bibinfo {author} {\bibfnamefont {R.}~\bibnamefont
  {Spreeuw}},\ and\ \bibinfo {author} {\bibfnamefont {J.}~\bibnamefont
  {Woerdman}},\ }\bibfield  {title} {\bibinfo {title} {Orbital angular momentum
  of light and the transformation of laguerre-gaussian laser modes},\
  }\href@noop {} {\bibfield  {journal} {\bibinfo  {journal} {Physical review
  A}\ }\textbf {\bibinfo {volume} {45}},\ \bibinfo {pages} {8185} (\bibinfo
  {year} {1992})}\BibitemShut {NoStop}%
\bibitem [{\citenamefont {Giovannini}\ \emph {et~al.}(2013)\citenamefont
  {Giovannini}, \citenamefont {Romero}, \citenamefont {Leach}, \citenamefont
  {Dudley}, \citenamefont {Forbes},\ and\ \citenamefont
  {Padgett}}]{Giovannini2013}%
  \BibitemOpen
  \bibfield  {author} {\bibinfo {author} {\bibfnamefont {D.}~\bibnamefont
  {Giovannini}}, \bibinfo {author} {\bibfnamefont {J.}~\bibnamefont {Romero}},
  \bibinfo {author} {\bibfnamefont {J.}~\bibnamefont {Leach}}, \bibinfo
  {author} {\bibfnamefont {A.}~\bibnamefont {Dudley}}, \bibinfo {author}
  {\bibfnamefont {A.}~\bibnamefont {Forbes}},\ and\ \bibinfo {author}
  {\bibfnamefont {M.~J.}\ \bibnamefont {Padgett}},\ }\bibfield  {title}
  {\bibinfo {title} {Characterization of high-dimensional entangled systems via
  mutually unbiased measurements},\ }\href@noop {} {\bibfield  {journal}
  {\bibinfo  {journal} {Phys. Rev. Lett.}\ }\textbf {\bibinfo {volume} {110}},\
  \bibinfo {pages} {143601} (\bibinfo {year} {2013})}\BibitemShut {NoStop}%
\bibitem [{\citenamefont {D'Ambrosio}\ \emph {et~al.}(2013)\citenamefont
  {D'Ambrosio}, \citenamefont {Cardano}, \citenamefont {Karimi}, \citenamefont
  {Nagali}, \citenamefont {Santamato}, \citenamefont {Marrucci},\ and\
  \citenamefont {Sciarrino}}]{DAmbrosio2013}%
  \BibitemOpen
  \bibfield  {author} {\bibinfo {author} {\bibfnamefont {V.}~\bibnamefont
  {D'Ambrosio}}, \bibinfo {author} {\bibfnamefont {F.}~\bibnamefont {Cardano}},
  \bibinfo {author} {\bibfnamefont {E.}~\bibnamefont {Karimi}}, \bibinfo
  {author} {\bibfnamefont {E.}~\bibnamefont {Nagali}}, \bibinfo {author}
  {\bibfnamefont {E.}~\bibnamefont {Santamato}}, \bibinfo {author}
  {\bibfnamefont {L.}~\bibnamefont {Marrucci}},\ and\ \bibinfo {author}
  {\bibfnamefont {F.}~\bibnamefont {Sciarrino}},\ }\bibfield  {title} {\bibinfo
  {title} {Test of mutually unbiased bases for six-dimensional photonic quantum
  systems},\ }\href {https://doi.org/10.1038/srep02726} {\bibfield  {journal}
  {\bibinfo  {journal} {Scientific Reports}\ }\textbf {\bibinfo {volume} {3}},\
  \bibinfo {pages} {2726} (\bibinfo {year} {2013})}\BibitemShut {NoStop}%
\bibitem [{\citenamefont {Malik}\ \emph {et~al.}(2012)\citenamefont {Malik},
  \citenamefont {O'Sullivan}, \citenamefont {Rodenburg}, \citenamefont
  {Mirhosseini}, \citenamefont {Leach}, \citenamefont {Lavery}, \citenamefont
  {Padgett},\ and\ \citenamefont {Boyd}}]{Malik12}%
  \BibitemOpen
  \bibfield  {author} {\bibinfo {author} {\bibfnamefont {M.}~\bibnamefont
  {Malik}}, \bibinfo {author} {\bibfnamefont {M.}~\bibnamefont {O'Sullivan}},
  \bibinfo {author} {\bibfnamefont {B.}~\bibnamefont {Rodenburg}}, \bibinfo
  {author} {\bibfnamefont {M.}~\bibnamefont {Mirhosseini}}, \bibinfo {author}
  {\bibfnamefont {J.}~\bibnamefont {Leach}}, \bibinfo {author} {\bibfnamefont
  {M.~P.~J.}\ \bibnamefont {Lavery}}, \bibinfo {author} {\bibfnamefont {M.~J.}\
  \bibnamefont {Padgett}},\ and\ \bibinfo {author} {\bibfnamefont {R.~W.}\
  \bibnamefont {Boyd}},\ }\bibfield  {title} {\bibinfo {title} {Influence of
  atmospheric turbulence on optical communications using orbital angular
  momentum for encoding},\ }\href {https://doi.org/10.1364/OE.20.013195}
  {\bibfield  {journal} {\bibinfo  {journal} {Opt. Express}\ }\textbf {\bibinfo
  {volume} {20}},\ \bibinfo {pages} {13195} (\bibinfo {year}
  {2012})}\BibitemShut {NoStop}%
\bibitem [{\citenamefont {Bennett}\ and\ \citenamefont
  {Brassard}(1984)}]{BB84}%
  \BibitemOpen
  \bibfield  {author} {\bibinfo {author} {\bibfnamefont {C.~H.}\ \bibnamefont
  {Bennett}}\ and\ \bibinfo {author} {\bibfnamefont {G.}~\bibnamefont
  {Brassard}},\ }\bibfield  {title} {\bibinfo {title} {Quantum cryptography:
  Public key distribution and coin tossing},\ }in\ \href@noop {} {\emph
  {\bibinfo {booktitle} {Proceedings of IEEE International Conference on
  Computers Systems and Signal Processing}}}\ (\bibinfo {address} {Bangalore,
  India},\ \bibinfo {year} {1984})\ pp.\ \bibinfo {pages}
  {175--179}\BibitemShut {NoStop}%
\bibitem [{Note1()}]{Note1}%
  \BibitemOpen
  \bibinfo {note} {We note here that in general, a projective measurement does
  not achieve the minimum possible guesswork. To do so, one would need to
  optimize over all quantum measurements, not just projective
  ones.}\BibitemShut {Stop}%
\bibitem [{\citenamefont {Ando}\ \emph {et~al.}(2009)\citenamefont {Ando},
  \citenamefont {Ohtake}, \citenamefont {Matsumoto}, \citenamefont {Inoue},\
  and\ \citenamefont {Fukuchi}}]{Ando09}%
  \BibitemOpen
  \bibfield  {author} {\bibinfo {author} {\bibfnamefont {T.}~\bibnamefont
  {Ando}}, \bibinfo {author} {\bibfnamefont {Y.}~\bibnamefont {Ohtake}},
  \bibinfo {author} {\bibfnamefont {N.}~\bibnamefont {Matsumoto}}, \bibinfo
  {author} {\bibfnamefont {T.}~\bibnamefont {Inoue}},\ and\ \bibinfo {author}
  {\bibfnamefont {N.}~\bibnamefont {Fukuchi}},\ }\bibfield  {title} {\bibinfo
  {title} {Mode purities of laguerre--gaussian beams generated via
  complex-amplitudemodulation using phase-only spatial light modulators},\
  }\href {https://doi.org/10.1364/OL.34.000034} {\bibfield  {journal} {\bibinfo
   {journal} {Opt. Lett.}\ }\textbf {\bibinfo {volume} {34}},\ \bibinfo {pages}
  {34} (\bibinfo {year} {2009})}\BibitemShut {NoStop}%
\end{thebibliography}%

\onecolumngrid

\renewcommand{\thefigure}{S\arabic{figure}}
\setcounter{figure}{0}
\renewcommand{\thetable}{S\arabic{table}}
\setcounter{table}{0}
\renewcommand{\theequation}{S.\arabic{equation}}
\setcounter{equation}{0}

\section*{Supplementary material}
In this Supplementary Material, we present additional information regarding the calculation of guesswork with quantum side information. In particular, we provide complete details regarding the analytical formula for the guesswork when a standard basis measurement is performed, and then we detail, with the help of an example, how the guesswork is calculated for the optimal projective measurement.

\appendix

\section{Guesswork calculation with a standard basis measurement}

\subsection{$d=2$}

In the main text, we stated that when the side information consists of the four BB84 states, the average number of guesses is $1.75$ when bob performs a standard basis ($\{ \ket{0}, \ket{1} \}$) measurement.

This number can be understood as follows: if the measurement outcome is $\ket{0}$, then the posterior probabilities of states $\{ \ket{0}, \ket{1}, \ket{+}, \ket{-} \}$ are $\{ 1/2, 0, 1/4, 1/4 \}$. Likewise, for outcome $\ket{1}$, the posterior probabilities are $\{ 0, 1/2, 1/4, 1/4 \}$. For outcome $\ket{0}$, Bob guesses in the order $(x_1, x_3, x_4, x_2)$ and for outcome $\ket{1}$, the guessing order is $(x_2, x_4, x_3, x_1)$. In each of these cases, the average number of guesses is $1 \cdot 1/2 + 2 \cdot 1/4 + 3 \cdot 1/4$ and thus the guesswork $G(X|B)_{\rho} = 1.75$.

\subsection{General dimension $d$}

In the main text, we showed that for the higher dimensional generalization of the BB84 example, the guesswork when using a standard basis measurement is ${(d+5)}/{4}$ for dimension $d \geq 2$. Here we show how this is obtained.

We recall that, for dimension $d$, the $2d$ side information states are divided into two mutually unbiased bases of $d$ states each:
\begin{equation}
	\{ \ket{0}, \ket{1}, \dots, \ket{d-1}, \ket{\widetilde{0}}, \ket{\widetilde{1}}, \dots, \ket{\widetilde{d-1}} \}.
\end{equation}

When Bob performs a standard basis measurement, it means that he projects his state onto the basis $\{ \ket{0}, \ket{1}, \dots, \ket{d-1} \}$. We know from Massey's criterion for optimizing the guesswork that Bob should guess in decreasing order of the posterior distribution of measurement outcomes. When outcome $\ket{k}$ is obtained, Bob can eliminate the $d-1$ computational basis states that are orthogonal to $\ket{k}$.

Given measurement outcome corresponding to $\ket{k}$, the posterior probabilities of the states $\{ \ket{0}, \ket{1}, \dots, \ket{k}, \dots, \ket{d-1}, \ket{\widetilde{0}}, \ket{\widetilde{1}}, \dots, \ket{\widetilde{d-1}} \}$ are $\{ 0, 0, \dots, 1/2, \dots, 0, 1/2d, 1/2d, \dots, 1/2d  \}$. This means that Bob's guessing order is $(x_k, x_{\widetilde{0}}, \dots, x_{\widetilde{d-1}}, x_0, \dots x_{k-1}, x_{k+1}, \dots x_{d-1})$. We note here that this guessing order is quite flexible--all that is required is that the first guess by $x_k$, the next $d$ guesses correspond to $x_{\widetilde{0}}$ through $x_{\widetilde{d-1}}$ in any order, and that the remaining standard basis symbols follow again in any order.

The probability of Bob's first guess being right is $1/2$, which we read off the posterior distribution. In case it is incorrect, then the probability of correctness of each of his next $d$ guesses is $1/2d$. We can use this to compute the expected value of the number of guesses:
\begin{align}
	\mathbb{E}[G(X|B)] &= 1 \cdot 1/2 + 2 \cdot 1/2d + \dots + (d + 1) \cdot 1/2d + 0 + \dots + 0 \\
						&= \frac{1}{2} + \frac{1}{2d} \left( 2 + 3 + \dots + (d+1) \right) \\
						&= \frac{1}{2} + \frac{1}{2d} \left( \frac{(d+1)(d+2)}{2} - 1 \right) \\
						&= \frac{1}{4d} (2d) + \frac{1}{4d} \left(d^2 + 3d \right) \\
						&= \frac{d + 5}{4}. \label{eq:std-basis-formula}
\end{align}

\section{Guesswork calculation with the optimal projective measurement}

First we detail how we optimize over all projective measurements for arbitrary dimension $d$. To do so, we consider the basis $\{ \ket{\psi_0}, \dots, \ket{\psi_{d-1}} \}$ to arise from a parameterized $d \times d$ unitary matrix. The matrix's rows correspond to the coefficients of each state of the orthonormal basis. The guessing orders for each measurement outcome are obtained by calculating the overlaps of each basis state with the generalized BB84 states and then arranging the generalized BB84 states in descending order of their overlaps, just like we described for the standard basis measurement in the preceding paragraph. The optimization procedure to determine the optimal measurement is straightforward, and is performed using the Global Optimization Toolbox in MATLAB. As an illustrative example, we describe the details of this calculation for $d=3$ below.

\subsection{Guesswork calculation for $d=3$ with the numerically determined measurement} \label{app:guesswork-example-calculation}

In Table I in the main text, we stated that the guesswork for the $d=3$ BB84 example was $2$ when using the standard basis measurement. This follows the ${(d+5)}/{4}$ formula that we derived just above in~\eqref{eq:std-basis-formula}. We also state that using a numerically predetermined projective measurement, we can do better and attain a value of $1.9425$. Here we show how this is achieved.

We recall that the six side information states are
\begin{equation}
	\{ \ket{0}, \ket{1}, \ket{2}, \ket{\widetilde{0}}, \ket{\widetilde{1}}, \ket{\widetilde{2}} \} \label{app:eq:side-info-states}
\end{equation} 
whose corresponding classical symbols are $x_0, x_1, x_2, x_{\widetilde{0}}, x_{\widetilde{1}} \text{ and } x_{\widetilde{2}}$.

Our numerical goal is to optimize overall possible projective measurements with the aim of minimizing the guesswork. Thus, we consider an arbitrary orthonormal basis $\{ \ket{\psi_0}, \ket{\psi_1}, \ket{\psi_{2}} \}$ onto which we project.

Given an arbitrary basis $\{ \ket{\psi_0}, \ket{\psi_1}, \ket{\psi_{2}} \}$, the first step is to calculate the overlaps of each of the six side information states \eqref{app:eq:side-info-states} with the three states given above. For each of $\ket{\psi_0}$, $\ket{\psi_1}$ and $\ket{\psi_2}$, we arrange the side information states in decreasing order of overlap. This gives us the optimal guessing order corresponding to each measurement outcome, and the overlaps for each measurement outcome yield the posterior distribution of the correct answer. Thus, the average number of guesses for each measurement outcome can be calculated. 

Further, Bob's state alone, before any measurement is done, is the maximally mixed state. This can be seen by considering the states in~\eqref{app:eq:side-info-states} and taking a uniformly distributed mixture of them. This means that each measurement outcome, averaged over input states, is equally likely. The overall guesswork is thus the average of the number of guesses for each measurement outcome.

So far, we have described how to calculate the guesswork corresponding to an arbitrarily chosen measurement basis $\{ \ket{\psi_0}, \ket{\psi_1}, \ket{\psi_{2}} \}$. All that remains is to optimize over all such measurement bases. These states are parameterized by considering a parameterization of a $d \times d$ unitary matrix given by Hedemann.  A $3 \times 3$ unitary matrix is parameterized by 6 complex parameters, $a$ through $f$, in the following manner:
\begin{equation}%
	U  = \left( {\begin{array}{*{20}c}
			a & {bc} & {bd}  \\
			{b^* e} & { - a^* ce - d^* f^* } & { - a^* de + c^* f^* }  \\
			{b^* f} & { - a^* cf + d^* e^* } & { - a^* df - c^* e^* }  \\
	\end{array}} \right),
\end{equation}
subject to the constraints $|a|^2+|b|^2 =1$, $|c|^2+|d|^2 =1$, and $|e|^2+|f|^2 =1$.

We thus have that
\begin{equation}
\begin{aligned} 
	\ket{\psi_0} &= a \ket{0} + bc \ket{1} + bd \ket{2} \\
	\ket{\psi_1} &= b^*e \ket{0} + (-a^*ce - d^*f^*) \ket{1} + (-a^*de + c^*f^*) \ket{2} \\
	\ket{\psi_{2}} &= b^* f \ket{0} + (- a^* cf + d^* e^*) \ket{1} +  (- a^* df - c^* e^* ) \ket{2}. 
\end{aligned} \label{app:d3-parameterized-states}
\end{equation}

We use MATLAB's Global Optimization Toolbox to optimize over the six variables and three constraints provided above, and calculate the optimal guesswork for this example across all projective measurements. The optimal states are given by the following coefficients:

\begin{align}
	\ket{\psi_0} &= \left( 0.7413 - 0.6421i, \quad 0.0118 + 0.1221i, \quad  -0.1085 - 0.1067i \right) \\
	\ket{\psi_1} &=  \left( -0.0919 + 0.1244i, \quad 0.0688 - 0.0060i, \quad  -0.8069 - 0.5659i \right)\\
	\ket{\psi_{2}} &= \left( 0.0676 + 0.0985i, \quad  0.9847 + 0.1023i, \quad  0.0634 + 0.0389i \right).
\end{align}

To calculate the guesswork from this, we construct the overlap table between the side information states and the measurement states as follows: \\

\begin{center}
	\begin{tabular}{|l|l|l|l|}
		\hline
		& $\psi_0$ & $\psi_1$ & $\psi_2$ \\ \hline
		$\ket{0}$             & 0.4809   & 0.0120   & 0.0071   \\ \hline
		$\ket{1}$             & 0.0075   & 0.0024   & 0.4901   \\ \hline
		$\ket{2}$             & 0.0116   & 0.4857   & 0.0028   \\ \hline
		$\ket{\widetilde{0}}$ & 0.2574   & 0.1170   & 0.1257   \\ \hline
		$\ket{\widetilde{1}}$ & 0.1347   & 0.1482   & 0.2171   \\ \hline
		$\ket{\widetilde{2}}$ & 0.1079   & 0.2348   & 0.1572   \\ \hline
	\end{tabular}
\end{center}

For each measurement outcome, the number of guesses can be calculated by arranging the corresponding column in descending order, and then taking an inner product of the column with $\left(1 2 3 4 5 6\right)$. We then get that the average number of guesses for each of the measurement outcomes is 1.9344, 1.9423 and 1.951. Taking an average of these three numbers yields
\begin{equation}
	G(X|B) = 1.9425.
\end{equation} 

\subsection{Guesswork calculation for $d=4$ with the numerically determined measurement}

A similar procedure as above is followed for the case of $d=4$. Just like we listed out three different states for $d=3$ in \eqref{app:d3-parameterized-states}, we have for four dimensions:

\begin{equation}
\begin{aligned} 
	\ket{\psi_0} &= a  \ket{0} + b^* g  \ket{1} + b^* h^* j \ket{2} + b^* h^* k \ket{3} \\
	\ket{\psi_1} &= b c \ket{0} + (-a^* c g + d^* h l) \ket{1} + (-a^* c h^* j - d^* g^* j l + d^* k^* m^*) \ket{2} + (-a^* c h^* k - d^* g^* k l - d^* j^* m^*) \ket{3} \\
	\ket{\psi_{2}} &= b d e \ket{0} + (-a^* d e g - c^* e h l + f^* h m) \ket{1} +  (-a^* d e h^* j + c^* e g^* j l - c^* e k^* m^* - f^* g^* j m - f^* k^* l^*) \ket{2} \\
	&+ (-a^* d e h^* k + c^* e g^* k l + c^* e j^* m^* - f^* g^* k m + f^* j^* l^*) \ket{3} \\
	\ket{\psi_{3}} &= b d f \ket{0} + (-a^* d f g - c^* f h l - e^* h m) \ket{1} +  (-a^* d f h^* j + c^* f g^* j l - c^* f k^* m^* + e^* g^* j m + e^* k^* l^*) \ket{2} \\
	&+ (-a^* d f h^* k + c^* f g^* k l + c^* f j^* m^* + e^* g^* k m - e^* j^* l^*) \ket{3}. \\
\end{aligned} \label{app:d4-parameterized-states}
\end{equation}

We use the same optimization technique used for the $d=3$ case and obtain the final set of measurement states as follows:

\begin{align}
	\ket{\psi_0} &= \left( -0.1116 - 0.04115i, \quad -0.0015 + 0.0131i, \quad  0.1336 - 0.0510i, \quad 0.0069 + 0.9824i  \right) \\
	\ket{\psi_1} &=  \left( -0.6943 + 0.6951i, \quad 0.05941 + 0.1301i, \quad  -0.081 + 0.0104i, \quad -0.1084 - 0.0490i  \right)\\
	\ket{\psi_{2}} &= \left( -0.1347 + 0.0481i, \quad  0.0138 - 0.9824i, \quad  0.1107 + 0.0435i, \quad 0.0018 - 0.0130i \right) \\
	\ket{\psi_{3}} &= \left( -0.0104 + 0.0080i, \quad  -0.0484 + 0.1086i, \quad  0.6991 + 0.6902i, \quad 0.1298 - 0.0602i \right).
\end{align}

Following the same procedure for calculating the guesswork as we did for $d=3$, we get the guesswork in this case to be
\begin{equation}
    G(X|B) = 2.1429.
\end{equation}

\end{document}